\documentclass[aps,prl,twocolumn,groupedaddress,nofootinbib,superscriptaddress,notitlepage,showpacs,floatfix]{revtex4-1}
\usepackage{graphicx,graphics}
\usepackage{epsfig}
\usepackage{subfigure}
\usepackage{lipsum}
\usepackage{dcolumn}   % needed for some tables
\usepackage{array}
\usepackage{amsthm}
\usepackage{amsmath}
\usepackage{amssymb}
\usepackage{amsfonts}
\usepackage{mathrsfs} %curled math symbols
\usepackage{epsfig}
\usepackage{bm}% bold math
\usepackage{braket} %Dirac notation
\usepackage{enumitem}
\usepackage{url}
\usepackage[pdfstartview=FitH]{hyperref}
\usepackage{pifont}
\hypersetup{
    colorlinks=true,        % false: boxed links; true: colored links
    linkcolor=blue,              % color of internal links
    citecolor=blue,        % color of links to bibliography
    filecolor=blue,          % color of file links
    urlcolor=blue,              % color of external links
    runcolor=cyan
			}
\usepackage[pdftex]{color}
\usepackage{multirow}
\usepackage{tikz}
\usepackage{CJK}
\newcommand{\tr}{\mathrm{tr}}
\newcommand{\ud}{\mathrm{d}}                % differential symbol
\newcommand{\defeq}{:=}           % be defined as
\begin{document}
%=================================================================================================================================================
% Title, authors, affiliation, dates, PACS.
%=================================================================================================================================================
\begin{CJK*}{GB}{}
\title{Perturbative analysis of quantum fluctuation theorems in a driven open system}
%------------------------------------------------------------------------------------------
\author{Yi Peng}
	\affiliation{Institute of Physics, Chinese Academy of Sciences, Beijing 100190, China}
	\affiliation{University of Chinese Academy of Sciences, Beijing 100049, China}
%------------------------------------------------------------------------------------------
\author{Heng Fan}
	\email{hfan@iphy.ac.cn}
	\affiliation{Institute of Physics, Chinese Academy of Sciences, Beijing 100190, China}
	\affiliation{University of Chinese Academy of Sciences, Beijing 100049, China}
	\affiliation{Collaborative Innovation Center of Quantum Matter, Beijing 100190, China}
%------------------------------------------------------------------------------------------
\date{\today}
\eid{identifier}
%=================================================================================================================================================
% Abstract
%=================================================================================================================================================
\begin{abstract}
	We give a perturbative analysis of Crooks relation and Jarzynski equality in an arbitrary driven quantum system weakly coupled to a heat bath.
	Invoking no efficient Hamiltonian nor any restriction on the form of the coupling, we derive the first-order correction to Crooks relation
	and Jarzynski equality in terms of the interacting Hamiltonian. A Crooks type of relation about energy increment in the system and 
	nonvanishing heat expenditure to the heat bath up to the second order of interaction would be given by the way. Our results tell us that 
	deviation of the quantum fluctuation relations is mainly caused by energy trapped interaction rather than the heat flow to (or from) heat 
	bath. We also show a way of defining work in weak coupling regime which is consistent to the work definition in closed systems and thus deepen
	our understanding of work in quantum realm.
\end{abstract}
\maketitle
\end{CJK*}
%=================================================================================================================================================
% Main contents
%=================================================================================================================================================
%=================================================================================================================================================
% Introduction
%=================================================================================================================================================
\emph{Introduction.}---The great success of equilibrium statistical mechanics has both deepended our understanding of nature and helped us
	build modern technical architecture. Though, nonequilibrium phenomenon has caught attention of physicists for a long time.
	Important results in this direction such as linear response theory were already given in near-equilibrium regime about half a centry 
	ago~\cite{Kubo1957}. With the thriving of nanotechnology in which fluctuations are significant, development of nonequilibrium statistics has 
	become more urgent. It stimulated a lot of investigation of statistical properties in systems arbitrarily out of equilibrium beyond linear 
	response theory. Fluctuation theorems are among the most remarkable achievements in the field of nonequilibrium statistical 
	mechanics~\cite{Bochkov1977,Jarzynski1997b,Crooks1999,Esposito2009,Campisi2011b}. Among them, Crooks relation and Jarzynski equality are of
	particular importance. Both theorems were first derived in a classical system weakly coupled to a heat bath of constant temperature 
	$T$~\cite{Jarzynski1997b,Crooks1999}. The former establishes a connection between a transition process and its reverse 
	\begin{equation}
		p\lbrack{W;\lambda}\rbrack / p\lbrack{-W;\tilde{\lambda}}\rbrack
		=e^{\beta\left(W-\Delta{F}_\mathrm{S}\right)}.
		\label{Crooks_relation}
	\end{equation}
	$p\lbrack{W;\lambda}\rbrack$ is the probability that $W$ work would be done on the system S which is driven out of equilibrium through
	an arbitrary protocol $\lambda$ while $p\lbrack{-W;\tilde{\lambda}}\rbrack$ probability of extracting that much of work from S during
	the mirrored reverse process. $\beta{\defeq}1/T$ denotes the inverse temperature and $\Delta{F}_\mathrm{S}$ the change of Helmholtz free 
	energy of system S. Notice that we adopt the convention of setting Boltzmann constant to one. From Crooks relation~(\ref{Crooks_relation}), 
	one can derive Jarzynski equality directly by rearranging (\ref{Crooks_relation}) and integrating over $W$
	\begin{equation}
		\Braket{e^{-\beta{W}}}_\lambda = e^{-\beta\Delta{F}_\mathrm{S}}.
		\label{Jarzynski_Equality}
	\end{equation}
	$\Braket{\bullet}_\lambda$ means averaging over statistical realizations of the forward process characterized by protocol $\lambda$. Many 
	attempts were made to extent them to the quantum regime since then~\cite{Esposito2009,Tasaki2000,Kurchan2000,Yukawa2000, Talkner2007a,
	Liu2014,Mukamel2003,DeRoeck2004,Esposito2006,Campisi2011b,Chetrite2012,Mehboudi2017,Talkner2009,Campisi2009,Campisi2009a,Esposito2009}. 
	The first task to achieve this goal is defining work in the framework of quantum mechanics which is quite delicate and has caused much 
	controversy~\cite{Talkner2007,Allahverdyan2014a,Jarzynski2015}. Among all these definition strategies, the two-point projective energy 
	measure (TPEM) protocol has been the most acceptable~\cite{Tasaki2000,Kurchan2000,Liu2014,Mukamel2003,DeRoeck2004,Esposito2006,Campisi2011b,
	Chetrite2012,Mehboudi2017,Talkner2009,Campisi2009,Campisi2009a,Esposito2009}. By using TPEM, quantum fluctuation theorems were first obtained 
	theoretically~\cite{Tasaki2000,Kurchan2000,Talkner2007a,Liu2014} and verified experimentally~\cite{An2014,Batalhao2014} for closed system 
	under unitary time evolution. Corresponding theoretical investigation for Markovian 
	processes~\cite{Mukamel2003,DeRoeck2004,Esposito2006,Chetrite2012,Mehboudi2017} were also developed. In the case where interaction with heat 
	bath is small and has vanishing expectation value in the eigenbasis of total Hamitonians at both ends of the transition, quantum fluctuation 
	theorems were also obtained~\cite{Talkner2009}. This covers many important physical models such as the Jaynes-Cummings 
	model~\cite{Shore1993} and the Rabi model~\cite{Braak2011,Xie2014} in the weak coupling regime. For interaction of other forms such as some 
	spin-spin coupling~\cite{Pfeuty1970,Quan2006} or that of sufficient magnitude, efficient Hamitonian of the system, consisting of free Hamiltonian of the
	system and an additional term resulting from interaction with heat bath, has to be employed to obtain Crooks relation and Jarzynski 
	equality~\cite{Talkner2009,Campisi2009,Campisi2009a}. 
	
	In this Letter, we consider also a quantum system  weakly coupled to a heat bath but with no further restriction on the form of interaction. 
	The system could be driven far away from equilibrium. We invoke no efficient Hamiltonian to incorporate the coupling to the heat bath and
	abandon the attempt of showing a formal consistency with the exact fluctuation relations as obtained in closed system and some specialized 
	open models. By invoking TPEM and proper definition of work, we examine Crooks~(\ref{Crooks_relation}) and 
	Jarzynski~(\ref{Jarzynski_Equality}) up to the first order of interaction. The work definition invoked in the weak coupling case can be proven
	to be consistent with that employed in a closed system~\cite{Tasaki2000,Kurchan2000,Esposito2009,Campisi2011b,Liu2014}. We will show with a 
	specific numerical example that without the efficient Hamiltonian, both relations would  break down and our correction would be effective to 
	some extend even when the interaction is comparable to the free Hamiltonian of system S. The correction turns
	out to be  the change of average coupling Hamiltonian. An energy fluctuation relation (\ref{CrooksRel_E_Q}) of Crooks type concerning energy 
	increment of system S and its nonvanishing heat expenditure to heat bath would be obtained up to the second order of interaction between 
	system S and heat bath B. In the weak coupling regime, energy trapped in interaction causes the deviation of Jarzynski equaltiy and Crook 
	relationn while impact of heat exchange with environment is minimal.
	
%=================================================================================================================================================
% System setup.
%=================================================================================================================================================
\emph{System setup.}---Consider the composite of system S and heat bath B
	\begin{equation}
		\pmb{H}\left(\lambda_t\right)
		= \pmb{H}_\mathrm{S}\left(\lambda_t\right) + \pmb{H}_\mathrm{B} + \pmb{H}_\mathrm{I},
	\end{equation}
	where coupling term $\pmb{H}_\mathrm{I}$ is a very small compared with $\pmb{H}_\mathrm{S}(\lambda_t)+\pmb{H}_\mathrm{B}$. $\lambda$ 
	represents a control protocol of a (set of) parameter(s) of S with $\lambda_t$ being its realization at time $t$. Thus, system Hamiltonian 
	$\pmb{H}_\mathrm{S}(\lambda_t)$ is time-dependent while heat bath Hamiltonian $\pmb{H}_\mathrm{B}$ and $\pmb{H}_\mathrm{I}$ are constant. The
	composite of S and B is initially a thermal state of temperature $T$ in both forward transition and its reverse
	\begin{eqnarray}
		\pmb{\rho}\left(\lambda_{0,\tau}\right)
		{\defeq} e^{-\beta\pmb{H}\left(\lambda_{0,\tau}\right)} / Z\left(\lambda_{0,\tau}\right).
	\end{eqnarray}
	Time duration of both the forward and backward transition is $\tau$. 
	$Z\left(\lambda_{0,\tau}\right){\defeq} \tr{e^{-\beta\pmb{H}\left(\lambda_{0,\tau}\right)}}$ are partition functions corresponding to 
	parameter settings $\lambda_0$ and $\lambda_\tau$ respectively.

%=================================================================================================================================================
% Driving and measurement protocol.
%=================================================================================================================================================
\emph{Measurements and driving protocol.}---System S and bath B start with a energy measurement which is the first part of TPEM at the begining of
	the forward transition
	\begin{equation}
		\check{\pmb{\rho}}\left(\lambda_0\right)
		{\simeq} \mathring{\pmb{\rho}}\left(\lambda_0\right)
			   \left(\openone-\beta\Delta\check{\pmb{H}}_\mathrm{I}^{\lambda_0}\right)
		\label{state_tot_post-1st-measurement_forward}
	\end{equation}
	where $\mathring{\pmb{\rho}}\left(\lambda_0\right)
	       {\defeq}\frac{\exp\{-\beta\left\lbrack{\pmb{H}_\mathrm{S}\left(\lambda_0\right)+\pmb{H}_\mathrm{B}}\right\rbrack\}}
						{Z_\mathrm{S}\left(\lambda_0\right)Z_\mathrm{B}}$ is free thermal state given a vanshing coupling 
	with partition functions 
	$Z_\mathrm{S}\left(\lambda_t\right){\defeq}\tr_\mathrm{S}e^{-\beta\pmb{H}_\mathrm{S}\left(\lambda_t\right)}$ for S at $t=0$ and
	$Z_\mathrm{B}{\defeq}\tr_\mathrm{B}e^{-\beta\pmb{H}_\mathrm{B}}$ for B. 
	The first-order correction term is 	
	$\Delta\check{\pmb{H}}_\mathrm{I}^{\lambda_0}
		\defeq \sum_{n,k}\left(\pmb{\Pi}_n^{\lambda_0}{\otimes}\pmb{\Pi}_k^\mathrm{B}\right)\pmb{H}_\mathrm{I}
						  \left(\pmb{\Pi}_n^{\lambda_0}{\otimes}\pmb{\Pi}_k^\mathrm{B}\right)
				-\Braket{\pmb{H}_\mathrm{I}}_{\lambda_0}.$	
	$\pmb{\Pi}_n^{\lambda_t}$ is the projective energy measurement of S at time $t$ such
	that $\pmb{H}\left(\lambda_t\right)\pmb{\Pi}_n^{\lambda_t} = E_n^{\lambda_t}\pmb{\Pi}_n^{\lambda_t}$ and similarly
	$\pmb{H}_\mathrm{B}\pmb{\Pi}_k^\mathrm{B}=E_k^\mathrm{B}\pmb{\Pi}_k^\mathrm{B}$ for heat bath. 
	$\braket{\bullet}_{\lambda_{0,\tau}}$ denotes average in a free thermal equilibrium state given specific configureation settings 
	$\lambda_{0,\tau}$ namely $\Braket{\pmb{H}_\mathrm{I}}_{\lambda_{0,\tau}}
	{\defeq}\tr_\mathrm{S}\lbrack{\mathring{\pmb{\rho}}(\lambda_{0,\tau})\pmb{H}_\mathrm{I}}\rbrack$,  while $\braket{\bullet}_{\lambda}$ is an 
	average evaluated over a process characterized by $\lambda$. Note that ``$\simeq$" means ``equal up to the first order of interacting 
	Hamiltonian $\pmb{H}_\mathrm{I}$".

	{\noindent}After the first phase of TPEM, system S and bath B would evolve unitary according to $\pmb{H}\left(\lambda_t\right)$ under the 
	driving protocol $\lambda$
	\begin{eqnarray}
		\pmb{U}_{\tau,0}\left\lbrack{\lambda}\right\rbrack
		\simeq\mathring{\pmb{U}}_{\tau,0}\left\lbrack{\lambda}\right\rbrack + \pmb{V}\lbrack{\lambda}\rbrack
		\label{UnitaryEvol_1st-OrderAppro}
	\end{eqnarray}
	where $\mathring{\pmb{U}}_{\tau,0}\left\lbrack{\lambda}\right\rbrack
			=\pmb{U}_{\mathrm{S};\tau,0}\lbrack{\lambda}\rbrack
	         {\otimes}\exp\left(-i\pmb{H}_\mathrm{B}\tau\right)$ describes the independent evolution of S and B when interaction is absent
	and $\pmb{V}\lbrack{\lambda}\rbrack$ is the first-order correction to $\pmb{U}_{\tau,0}\left\lbrack{\lambda}\right\rbrack$ in terms of the 
	coupling $\pmb{H}_\mathrm{I}$. If isolated from heat bath, the evolution of S would be governed by unitary time transition 
	$\pmb{U}_{\mathrm{S};\tau,0}\lbrack{\lambda}\rbrack{\defeq}
	\mathcal{T}\exp\left\lbrack{-i\int_0^\tau\ud{t}\pmb{H}_\mathrm{S}\left(\lambda_t\right)}\right\rbrack$. $\mathcal{T}$ is the 
	so-called time-ordering operator. Note we chose the convension of $\hbar=1$. Measurements of total energy 
	$\pmb{\Pi}_n^{\lambda_\tau}\otimes\pmb{\Pi}_k^\mathrm{B}$ as the second step of TPEM would be deployed at the end of the driving protocol. On
	the other hand, the reverse process would start from $\pmb{\rho}(\tilde{\lambda}_0)$, followed by measurement 
	$\pmb{\Pi}_n^{\tilde{\lambda}_0}\otimes\pmb{\Pi}_k^\mathrm{B}$, total unitary evolution $\pmb{U}_{\tau,0}\lbrack{\tilde{\lambda}}\rbrack$ and
	finally measurement $\pmb{\Pi}_n^{\tilde{\lambda}_\tau}\otimes\pmb{\Pi}_k^\mathrm{B}$ where $\tilde{\lambda}$ is the reverse protocol to 
	$\lambda$ such that $\tilde{\lambda}_{t}=\lambda_{\tau-t}$.

%=================================================================================================================================================
% Fluctuations of energy absorption and heat transfer.
%=================================================================================================================================================
\emph{Fluctuations of energy increment and heat transfer.}---In the forward process, probability of system energy increasing by $E$ while
	expending heat $Q$ to heat bath should be
	\begin{widetext}
		\begin{equation}
			p\left\lbrack{E,Q;\lambda}\right\rbrack
			= \sum_{n,k,m,\ell}\delta\left(E-E^{\lambda_\tau}_n+E^{\lambda_0}_{m}\right)\delta\left(Q-E^\mathrm{B}_k+E^\mathrm{B}_{\ell}\right)
			                   p\lbrack{m\ell,{nk};\lambda}\rbrack.
			\label{prob_E_Q_forward}
		\end{equation}
	\end{widetext}
	Probability $p\lbrack{m\ell,{nk};\lambda}\rbrack$ of obtaining $m$ and $\ell$ in the first projective energy measurement while $n$ and $k$ 
	in the second, should be 
	$\tr\{\left(\pmb{\Pi}_n^{\lambda_\tau}{\otimes}\pmb{\Pi}_k^\mathrm{B}\right)
											 \pmb{U}_{\tau,0}\left\lbrack{\lambda}\right\rbrack
											 \left(\pmb{\Pi}_m^{\lambda_0}{\otimes}\pmb{\Pi}_\ell^\mathrm{B}\right)
											 \check{\pmb{\rho}}\left(\lambda_0\right)
											 \pmb{U}^\dag_{\tau,0}\left\lbrack{\lambda}\right\rbrack\}$.
	Probability distribution $p\lbrack{-E,-Q;\tilde{\lambda}}\rbrack$ of system energy decrement and heat absorption during the backward 
	process would be of similar form. We can show that probability $p\lbrack{m\ell,{nk};\lambda}\rbrack$ of non-zero heat transfer namely
	$k\neq\ell$, generally are of the second order of coupling Hamiltonian $\pmb{H}_\mathrm{I}$ 
	\begin{widetext}
		\begin{equation}
			p\lbrack{m\ell,{nk};\lambda}\rbrack \approxeq 
													\tr\left\lbrack{\left(\pmb{\Pi}_n^{\lambda_\tau}\otimes\pmb{\Pi}_k^\mathrm{B}\right)
														\pmb{V}\lbrack{\lambda}\rbrack
														\left(\pmb{\Pi}_m^{\lambda_0}\otimes\pmb{\Pi}_\ell^\mathrm{B}\right)
														\pmb{V}^\dag\lbrack{\lambda}\rbrack
													}\right\rbrack
													{e^{-\beta(E_m^{\lambda_0}+E_\ell^\mathrm{B})}}\big/
			                                        {Z_\mathrm{B}Z_\mathrm{S}(\lambda_0)}.
		\end{equation}
	\end{widetext}
	``$\approxeq$" means ``equal up to the second order of the interacting Hamiltonian $\pmb{H}_\mathrm{I}$". Similar result can be derived
	for probability $p\lbrack{nk,{m\ell};\tilde{\lambda}}\rbrack$ in backward process. A direct conseqence of the microscopic time 
	reversibility $\pmb{U}_{\tau,0}^\dag\lbrack\tilde{\lambda}\rbrack=\pmb{U}_{\tau,0}\lbrack{\lambda}\rbrack$ and 
	$\mathring{\pmb{U}}_{\tau,0}^\dag\lbrack\tilde{\lambda}\rbrack=\mathring{\pmb{U}}_{\tau,0}\lbrack{\lambda}\rbrack$ of the composite system
	for both coupled and decoupled cases, is $\pmb{V}\lbrack{\tilde{\lambda}}\rbrack=\pmb{V}^\dag\lbrack{\lambda}\rbrack$. We can furhter use this
	to derive a close relation of the former pair of probabilities 
	\begin{equation}
		p\lbrack{m\ell,{nk};\lambda}\rbrack 
		\approxeq p\lbrack{nk,{m\ell};\tilde{\lambda}}\rbrack
		 e^{\beta(E_n^{\lambda_\tau}-E_m^{\lambda_0}
		          +E_k^\mathrm{B}-E_\ell^\mathrm{B}-\Delta{F}_\mathrm{S})},
		\label{CrooksRel_E_Q_discrete}
	\end{equation}
	where $\Delta{F}_\mathrm{S}{\defeq}F_\mathrm{S}\left(\lambda_\tau\right)- F_\mathrm{S}\left(\lambda_0\right)$ is Helmholtz free energy 
	difference of system S with parameter settings at the two ends of the forward process 
	$F_\mathrm{S}\left(\lambda_{0,\tau}\right){\defeq}-T\ln{Z_\mathrm{S}(\lambda_{0,\tau})}$. This relation holds for any nonvanishing heat 
	transferation where $k\neq\ell$. By substituting (\ref{CrooksRel_E_Q_discrete}) to the expression (\ref{prob_E_Q_forward}), we can derive a 
	fluctuation relation of Crooks form for energy increment $E$ and non-zero heat transfer $Q{\neq}0$
	\begin{equation}
		p\lbrack{E,Q;\lambda}\rbrack/p\lbrack{-E,-Q;\tilde{\lambda}}\rbrack 
		\approxeq e^{\beta(E+Q-\Delta{F}_\mathrm{S})}.
		\label{CrooksRel_E_Q}
	\end{equation}

%=================================================================================================================================================
% Correction to Crooks relation and Jarzynski equality.
%=================================================================================================================================================
\emph{Correction to Crooks relation and Jarzynski equality.}---
	Consider the case where eigengy spectrum of system S is non-degeneration, or we can employ measurements of complete sets of mutually 
	compatible observables on S which including system Hamiltonian $\pmb{H}_\mathrm{S}(\lambda_{0,\tau})$ in TPEM. We can then find eigenket 
	$\ket{\psi_{m}^{\lambda_{0,\tau}}}$ of the set of complete set of mutually compatible observables containing 
	$\pmb{H}_\mathrm{S}(\lambda_{0,\tau})$ such that 
	$\pmb{H}_\mathrm{S}(\lambda_{0,\tau})\ket{\psi_{m}^{\lambda_{0,\tau}}}=E_n^{\lambda_{0,\tau}}\ket{\psi_{m}^{\lambda_{0,\tau}}}$, with other 
	degrees of freedom within the degeneracies of the energy levels incorporated to notation $m$. Given vanishing coupling, transition 
	probability from $\ket{\psi_{m}^{\lambda_0}}$ to $\ket{\psi_{n}^{\lambda_\tau}}$ would be $\mathring{p}\lbrack{n|m;\lambda}\rbrack{\defeq}
	|\braket{\psi_{n}^{\lambda_\tau}|\pmb{U}_{\mathrm{S};\tau,0}\lbrack{\lambda}\rbrack|\psi_{m}^{\lambda_0}}|^2$ in the forward 
	drinving $\lambda$. Plugging $\mathring{p}\lbrack{n|m;\lambda}\rbrack$, (\ref{state_tot_post-1st-measurement_forward}) and 
	(\ref{UnitaryEvol_1st-OrderAppro}) onto (\ref{prob_E_Q_forward}) and invoking microscopic time reversibility 
	$\pmb{U}_{\tau,0}^\dag\lbrack\tilde{\lambda}\rbrack=\pmb{U}_{\tau,0}\lbrack{\lambda}\rbrack$ of the composite system, we can show that 
	$p\left\lbrack{E,Q;\lambda}\right\rbrack$ and $p\lbrack{-E,-Q;\tilde{\lambda}}\rbrack$ are mutually related up to the first order of 
	interaction  
	\begin{widetext}
		\begin{eqnarray}
			&&Z_\mathrm{S}(\lambda_0)p\lbrack{E,Q;\lambda}\rbrack
			 -e^{\beta(E+Q)}Z_\mathrm{S}(\lambda_\tau)p\lbrack{-E,-Q;\tilde{\lambda}}\rbrack				\nonumber\\
			&\simeq&\beta\sum_{m,n}\delta\left(E-E_n^{\lambda_\tau}+E_m^{\lambda_0}\right)
										\delta\left(Q\right)
										e^{-\beta{E}^{\lambda_0}_m}\mathring{p}\lbrack{n|m;\lambda}\rbrack
					\Big\lbrack\left(\braket{\psi_{n}^{\lambda_\tau}|\pmb{H}_\mathrm{I}^\mathrm{eff}|\psi_{n}^{\lambda_\tau}}
												-\braket{\pmb{H}_\mathrm{I}}_{\lambda_\tau}
				   \right)
				  -\left(\braket{\psi_{m}^{\lambda_0}|\pmb{H}_\mathrm{I}^\mathrm{eff}\psi_{m}^{\lambda_0}}
											  -\braket{\pmb{H}_\mathrm{I}}_{\lambda_0}
				   \right)
			\Big\rbrack.
			\label{ProbOfE&Q_relation}
		\end{eqnarray}
	\end{widetext} 	
	$\pmb{H}_\mathrm{I}^\mathrm{eff}{\defeq}\tr_\mathrm{B}\left(\pmb{H}_\mathrm{I}e^{\beta\pmb{H}_\mathrm{B}}/Z_\mathrm{B}\right)$ can be 
	considered as an effective potential applied to system S due to its interaction with heat bath B. 

	{\noindent}$\pmb{H}_\mathrm{I}$ generally would not commute with free Hamiltonian $\pmb{H}_\mathrm{S}(\lambda_{0,\tau})+\pmb{H}_\mathrm{B}$ 
	and thus they cannot be measured simultaneously. Averagely in a realization obtaining obtains $m$ and $\ell$ in the first measurement 
	while $n$ and $k$ in the second, the first law of thermodynamics is
	\begin{equation}
		W = E_n^{\lambda_\tau} - E_m^{\lambda_0} + E_k^\mathrm{B} - E_\ell^\mathrm{B} 
			+\Delta{E}_{nk,m\ell}^{\mathrm{I};\lbrack{\lambda}\rbrack}.
		\label{1st_law}
	\end{equation}
	$W$ is the work done by the external agent through the driving protocol 
	$\lambda$. $\Delta{E}_{nk,m\ell}^{\mathrm{I};\lbrack{\lambda}\rbrack}$ denotes change of the average coupling energy in the realization. 
	We can also have the first law of thermodynamics in the backward process with a similar definition given to the corresponding average change 
	of coupling energy $\Delta{E}_{m\ell,nk}^{\mathrm{I};\lbrack{\tilde{\lambda}}\rbrack}$. We will prove the validity of  the first law 
	(\ref{1st_law}) employed here later.

	%==================================================================================================================================
	\begin{figure*}[th!]
		\subfigure[\label{EQCrooks_Diference}$D_0(W_0,\omega_\mathrm{c})$]{\includegraphics[scale=0.33]{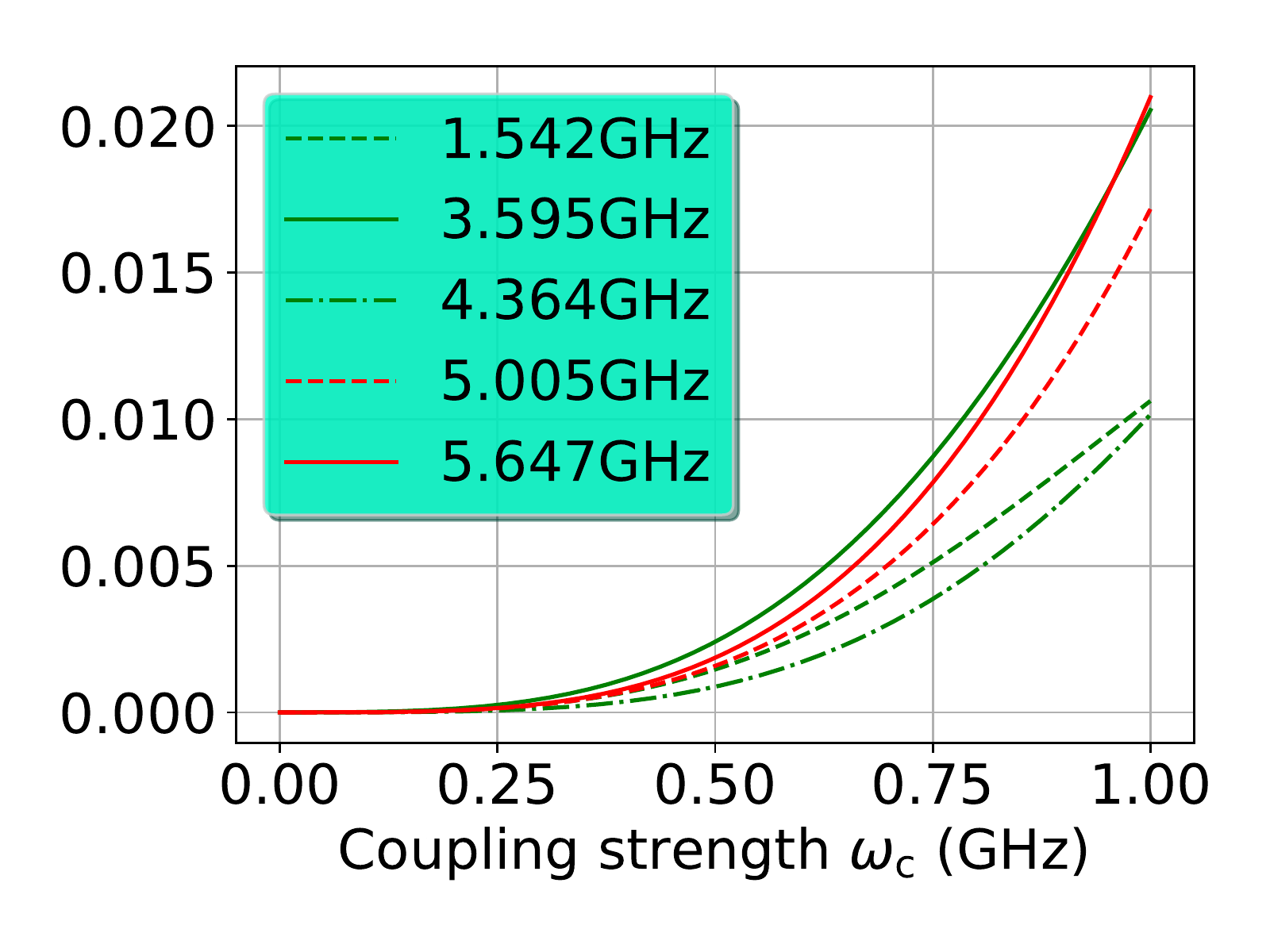}}
		\subfigure[\label{EQCrooks_Diference_derivative}$\partial{D_0(W_0,\omega_\mathrm{c})}/{\partial{\omega_\mathrm{c}}}$]
		{\includegraphics[scale=0.33]{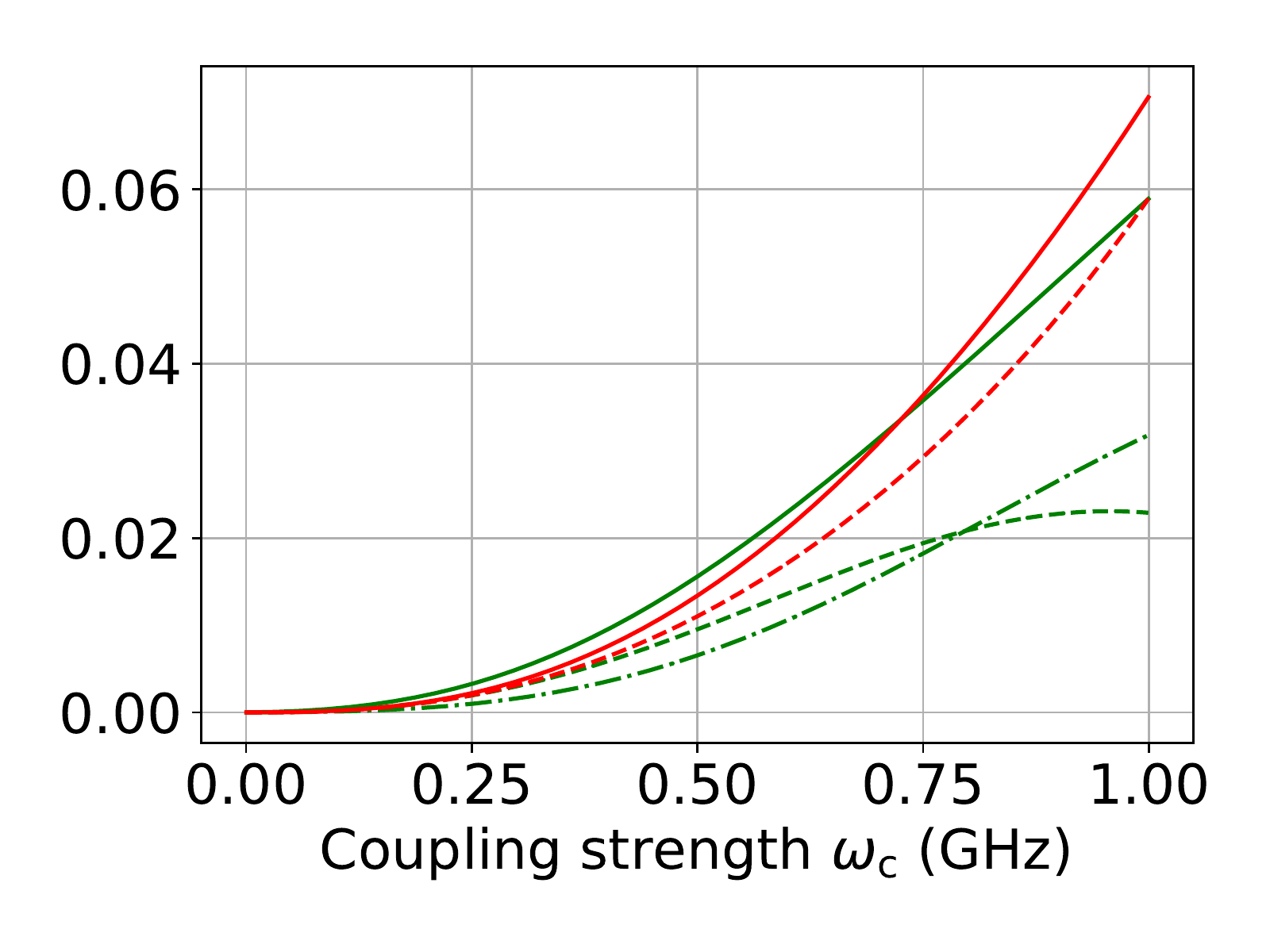}}
		\subfigure[\label{EQCrooks_differentia_prob}$\partial\{\int_{Q\neq0}\ud{Q}p\lbrack{W_0-Q,Q;\lambda}\rbrack\}/\partial{\omega_\mathrm{c}}$]
		{\includegraphics[scale=0.33]{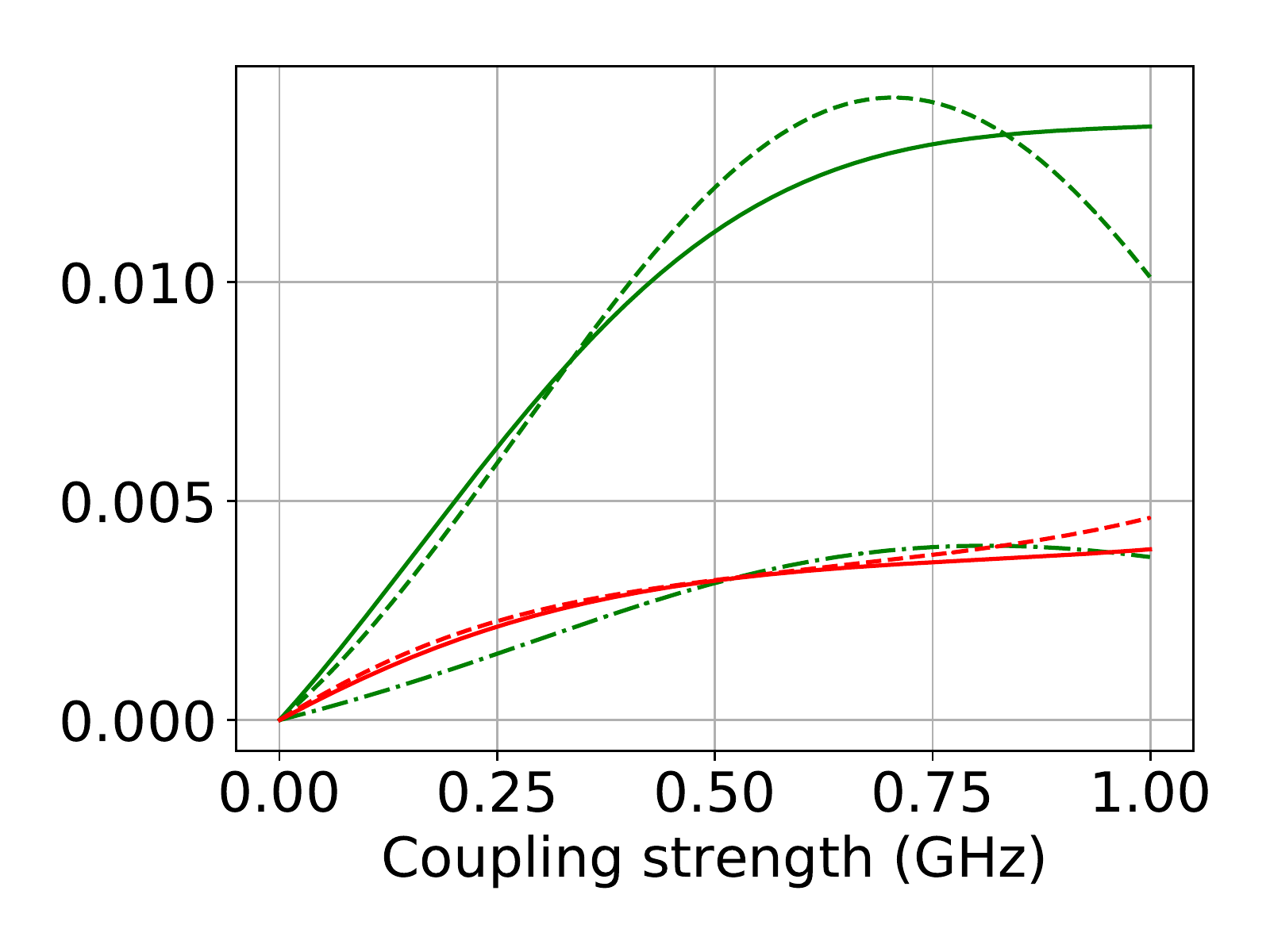}}
		\caption{(Colored online.)  $W_0$ with $|D_0(W_0,\omega_c=\omega)|\ge{0.01}$ are picked out as listed on the insets of both (a). The lines
				in (b) and (c) correspond to the same $W_0$ as the lines in (a) of the same color and line type as listed on the inset on the
				left hand. (c) tells us that probability of nonvanishing heat transferation does not vanishing in terms of the second order of
				coupling though its zeroth and first order do.}
		\label{EQCrooks}
	\end{figure*}
	%==================================================================================================================================
	{\noindent}According to the first law~(\ref{1st_law}), probability of injecting work $W$ to system S in the forward process would be
	\begin{widetext}
		\begin{eqnarray*}
			p\lbrack{W;\lambda}\rbrack
			&=&\int\ud{E}\ud{Q}\sum_{m,n,k,\ell}\delta\left(W-E-Q-\Delta{E}_{nk,m\ell}^{\mathrm{I};\lbrack{\lambda}\rbrack}\right)
											\delta\left(E-E_n^{\lambda_\tau}+E_m^{\lambda_0}\right)
											\delta\left(Q-E_k^\mathrm{B}+E_\ell^\mathrm{B}\right)
											p\lbrack{m\ell,{nk};\lambda}\rbrack,
		\end{eqnarray*}
	\end{widetext}
	From  energy fluctuation relation~(\ref{ProbOfE&Q_relation}), one can derive a similar relation between probability distributions of work 
	injection $p\lbrack{W;\lambda}\rbrack$ in forward process and extraction $p\lbrack{-W;\tilde{\lambda}}\rbrack$ in its reverse up to the first
	order of interaction
	\begin{widetext}
		\begin{equation}
			Z_\mathrm{S}(\lambda_0)p\lbrack{W;\lambda}\rbrack-e^{\beta{W}}Z_\mathrm{S}(\lambda_\tau)p\lbrack{-W;\tilde{\lambda}}\rbrack
			\simeq\beta\left(Z_\mathrm{S}(\lambda_0)p\lbrack{W;\lambda}\rbrack\braket{\pmb{H}_\mathrm{I}}_{\lambda_0}
									-e^{\beta{W}}Z_\mathrm{S}(\lambda_\tau)p\lbrack{-W;\tilde{\lambda}}\rbrack
									 \braket{\pmb{H}_\mathrm{I}}_{\lambda_\tau}
							  \right).
			\label{PobsOfwork_relation}
		\end{equation}
	\end{widetext}
	We can obtain the first-order correction to Crooks relation from (\ref{PobsOfwork_relation})
	\begin{equation}
		p\lbrack{W;\lambda}\rbrack/p\lbrack{-W;\tilde{\lambda}}\rbrack
		\simeq e^{\beta\left(W-\Delta{F}_\mathrm{S}-\Delta{E}_\mathrm{I}\right)}.
		\label{Crooks_like_relation}
	\end{equation}
	$\Delta{E}_\mathrm{I}{\defeq}\braket{\pmb{H}_\mathrm{I}}_{\lambda_\tau}-\braket{\pmb{H}_\mathrm{I}}_{\lambda_0}$ is difference of interaction
	energy of the compound system of S and B, given different parameter settings $\lambda_{0,\tau}$ of S. By appropriately rearrangement of
	(\ref{Crooks_like_relation}) and integration over $W$, one can derive the first-order correction of Jarzynski equality
	\begin{equation}
		\braket{e^{-\beta{W}}}_\lambda 
		\simeq e^{-\beta\left(\Delta{F}_\mathrm{S} +\Delta{E}_\mathrm{I}\right)}.
		\label{Jarzynski_like_relation}
	\end{equation}

%=================================================================================================================================================
%	Consistency between work definitions in the weak coupling and closed cases.
%=================================================================================================================================================
\emph{Consistency between work definitions in the weak coupling and closed cases.}
	If one consider the complex of system S and heat bath B as a whole, it is closed. Its partition functions corresponding to the two 
	configurations $\lambda_{0,\tau}$ of subsystem S should be 
	$Z(\lambda_{0,\tau}){\simeq}Z_\mathrm{S}(\lambda_{0,\tau})Z_\mathrm{B}(1-\beta\braket{\pmb{H}_\mathrm{I}}_{\lambda_{0,\tau}})$. Therefore,
	the change of Helmholtz free energy of the total system should be
	\begin{equation}
		\Delta{F}\simeq\Delta{F}_\mathrm{S}+\Delta{E}_\mathrm{I}.
		\label{F_tot_1st-order_correction}
	\end{equation}
	Hence corrected Crooks relation (\ref{Crooks_like_relation}) of subsystem S would coincide with the Crooks relation for the total system where
	TPEM is implemented for total Hamiltonian $\pmb{H}(\lambda_{0,\tau})$ instead of sum of free Hamiltonians of subsystems 
	$\pmb{H}_\mathrm{S}(\lambda_{0,\tau})+\pmb{H}_\mathrm{B}$ up the first order of $\pmb{H}_\mathrm{I}$. Analogical inference can be made to 
	corrected Jarzynski equality (\ref{Jarzynski_like_relation}) of S. Therefore, the work definition (\ref{1st_law}) namely the first law we 
	employed would conincide with those employed in closed systems~\cite{Tasaki2000,Kurchan2000,Talkner2007a,Liu2014} in weak coupling regime.

	%==================================================================================================================================
	\begin{figure*}[th!]
		\begin{center}
			\subfigure[\label{CR_1st_correction}$D(W,\omega_\mathrm{c})$]{\includegraphics[scale=0.33]{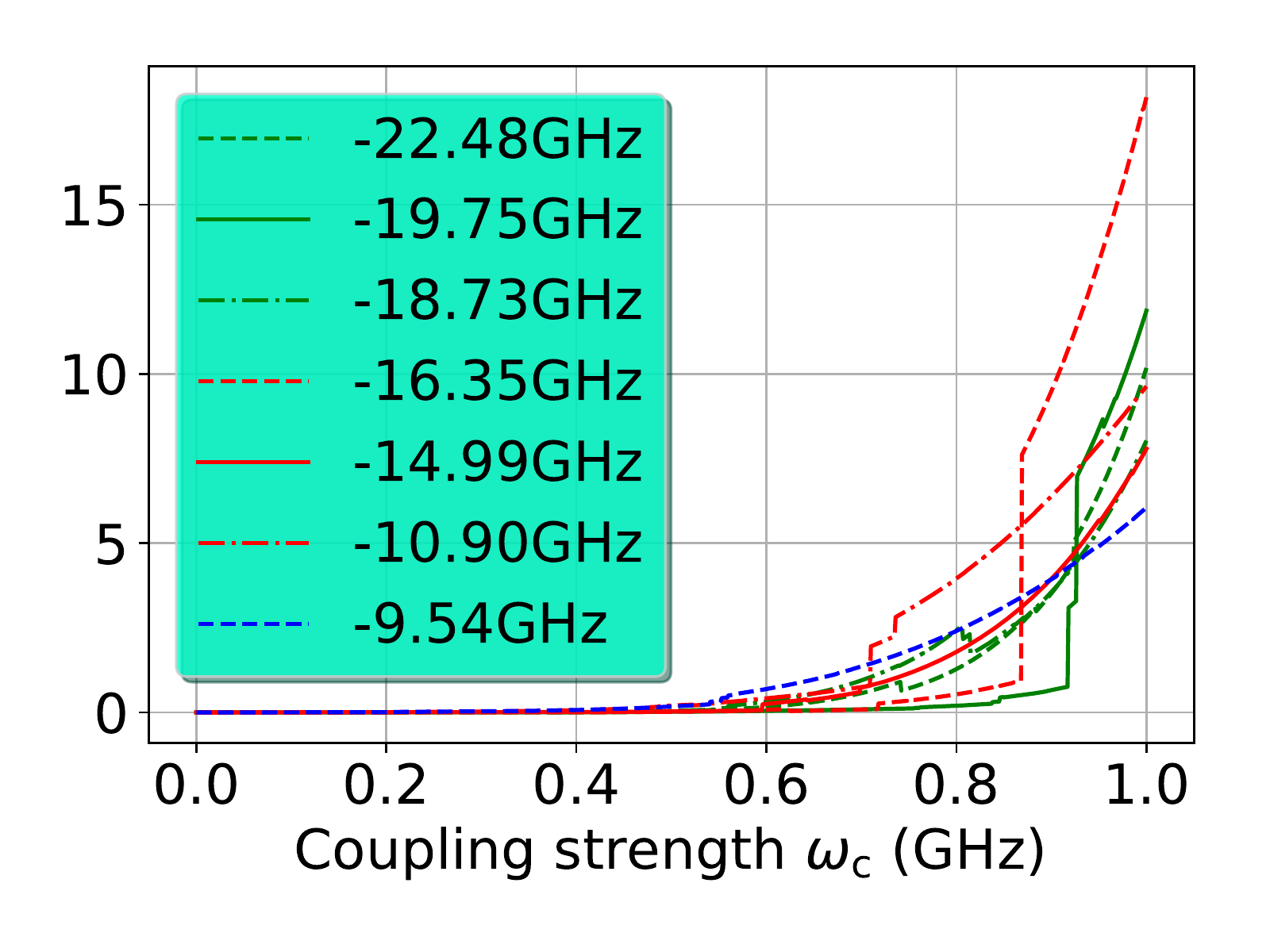}}
			\subfigure[\label{CR_1st_correction_dev_sig}$p\lbrack{W;\lambda}\rbrack$]
			          {\includegraphics[scale=0.33]{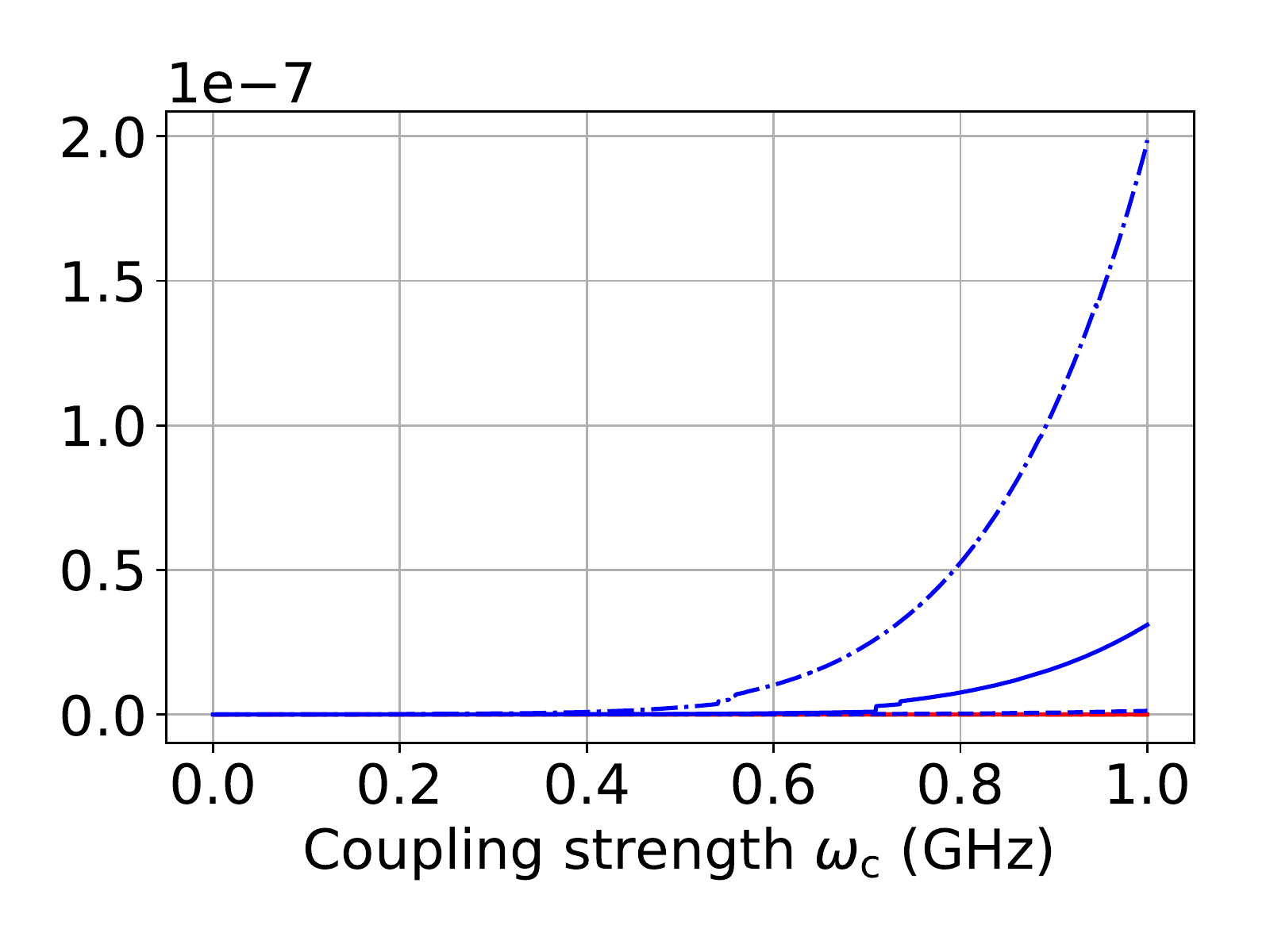}}
			\subfigure[\label{JE_correction}Correction to Jarzynski equality.]{\includegraphics[scale=0.33]{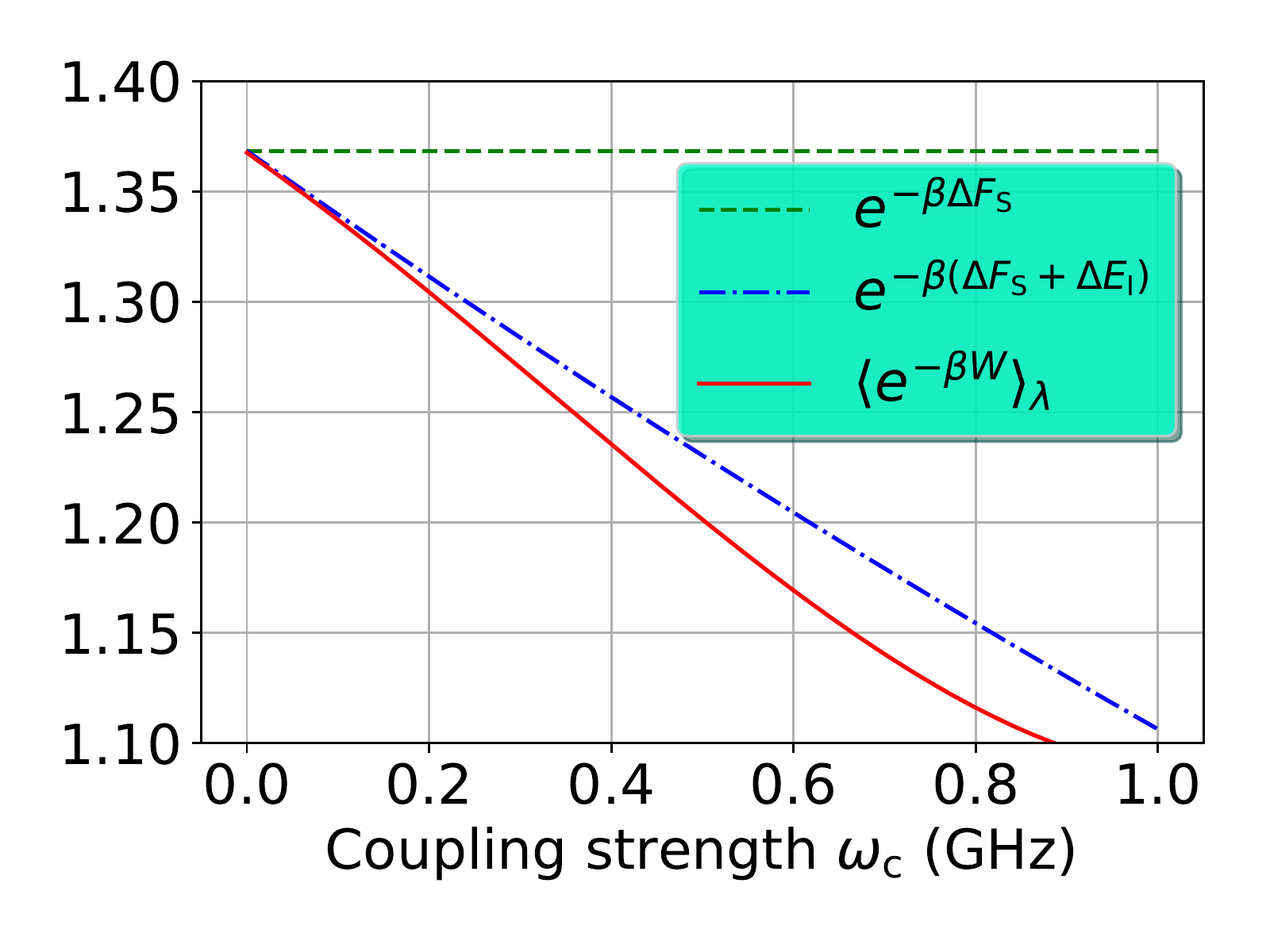}}
			\caption{(Colored online.)  $W$ with $|D(W,\omega_c=\omega)|\ge{4}$ are picked out as listed on the insets of (a). The lines
				in (b) correspond to the same $W$ as the lines in (a) of the same color and line type as listed on the inset on the
				left hand.}
		\end{center}
	\end{figure*}
	%==================================================================================================================================
%=================================================================================================================================================
% Specific model exemplification.
%=================================================================================================================================================
\emph{Specific model exemplification.}---By employing a two-level system (TLS) coupled to an  open-bounded transverse Ising chain (OTIC) of length
	$L=8$, we demonstrate fluctuation relation (\ref{CrooksRel_E_Q}) of energy increment of TLS and nonvanishing heat transfering to OTIC, and 
	show the effect of our correction (\ref{Crooks_like_relation}) to Crooks relation and (\ref{Jarzynski_like_relation}) to Jarzynski equation. 
	TLS is the system of interest corresponding to system S and OTIC acts as heat bath B. We assume Hamiltonian of TLS to be 
	$
	\pmb{H}_\mathrm{TLS}(t) {\defeq} (1+t/\tau)\omega\lbrack{\pmb{\sigma}_z\cos(\omega_\mathrm{prec}t)
																	 +\pmb{\sigma}_x\sin(\omega_\mathrm{prec}t)}\rbrack,
	$
	Hamitolnian of OTIC 
	$
		\pmb{H}_\mathrm{OTIC} {\defeq} -J\lbrack{\sum_{j=1}^{L-1}\pmb{\sigma}_z^{(j)}\pmb{\sigma}_z^{(j+1)}
							    +\gamma\sum_{i=1}^{L}\pmb{\sigma}_x^{(i)} }\rbrack,
	$
	and the coupling between them  
	$
		\pmb{H}_\mathrm{I}{\defeq} \omega_\mathrm{c}\ket{\uparrow_z}\bra{\uparrow_z}
		         \otimes\lbrack{\pmb{\sigma}_x^{(1)}+\pmb{\sigma}_x^{(L)}}\rbrack.
	$
	$\pmb{\sigma}_{x,y,z}^{(i)}$ are Pauli matrices belong to the $i$th site of OTIC while $\pmb{\sigma}_{x,y,z}$ are from TLS. 
	$\ket{\uparrow_z}$ is a state of TLS pointing up along the $z$ direction. We set $\omega=1~\mathrm{GHz}$, $J=2~\mathrm{GHz}$ and 
	$\gamma=1/2$. The precessing angular velocity is set to $\omega_\mathrm{prec}=0.5~\mathrm{GHz}$ and transition duration $\tau=\pi/\omega$. We
	assume temperature to be $7.64~\mathrm{mk}$. The coupling strength $\omega_\mathrm{c}$ varies from $0$ to $\omega$.

	{\noindent}Since it is much more involved to confront fluctuation relation (\ref{CrooksRel_E_Q}) directly, instead we choose to study 
	\begin{equation}
		\frac{\int_{Q\neq{0}}{\ud{Q}}p\lbrack{W_0-Q,Q;\lambda}\rbrack}{\int_{Q\neq{0}}{\ud{Q}}p\lbrack{-W_0+Q,-Q;\tilde{\lambda}}\rbrack}
		\approxeq e^{\beta(W_0-\Delta{F}_\mathrm{S})}
		\label{CrooksRel_E_Q_Variant}
	\end{equation}
	which is a direct consequence of (\ref{CrooksRel_E_Q}). For further simplification, we conduct coarse-grained division of the domain of
	$W_0$ by pieces of width $\delta{W_0}=0.13~\mathrm{GHz}$. For every $\omega_\mathrm{c}$ between $0$ and $\omega$, we numerically evluate the 
	difference function $D_0(W_0,\omega_c){\defeq}\int_{Q\neq{0}}\ud{Q}\{p\lbrack{-W_0+Q,-Q;\tilde{\lambda}}\rbrack{e^{\beta(W_0
	                                                                     -\Delta{F}_\mathrm{S})}}-p\lbrack{W_0-Q,Q;\lambda}\rbrack\}$ which would
	vanish up to the second order of small  coupling strength $\omega_\mathrm{proc}$ according to (\ref{CrooksRel_E_Q_Variant}). Our numerical
	simulation tells us that both $D(W_0,\omega_c)$ (c.f.  Fig.~\ref{EQCrooks_Diference}) and its derivative 
	(c.f. Fig.~\ref{EQCrooks_Diference_derivative}) vanishes in the vicinity of vanishing coupling. Thus give strong numerical evidence for
	(\ref{CrooksRel_E_Q_Variant}) and (\ref{CrooksRel_E_Q}) being valid in the second order of the coupling Hamiltonian. One can infer from
	Fig.~\ref{EQCrooks_differentia_prob} that (\ref{CrooksRel_E_Q_Variant}) and (\ref{CrooksRel_E_Q}) are nontrivial in general. We remark that
	driving protocols $\lambda$ and $\tilde{\lambda}$ contain information of $\omega_\mathrm{proc}$.

	{\noindent}Similarly, we use coarse-grained division of the domian of work $W$ by pieces of $\delta{W}=0.34~\mathrm{GHz}$ in the numerical 
	demonstration of the first-order correction of (\ref{Crooks_like_relation}) to Crooks relation. We evaluate the difference function 
	$D(W,\omega_\mathrm{c}){\defeq}p\lbrack{-W;\tilde{\lambda}}\rbrack{e^{-\beta{W}}}-p\lbrack{W;\lambda}\rbrack$ which would vanish in the first 
	order of $\omega_\mathrm{c}$ according to (\ref{Crooks_like_relation}). The simulation results shown in Fig.~\ref{CR_1st_correction} comfirms
	us the validity the correction (\ref{Crooks_like_relation}). From Fig.~\ref{CR_1st_correction_dev_sig}, we can tell that in strong coupling 
	regime $\omega_\mathrm{c}\ge{0.2~\mathrm{GHz}}$ the probability of enountering significant deviation 
	$|D(W,\omega_\mathrm{c}=\omega)|\ge{4}$ is very thin ($\sim10^{-7}$).
	Numerical results of first-order correction (\ref{Jarzynski_like_relation}) to Jarzynski equality are shown in 
	Fig.~\ref{JE_correction}. It tells us that our first-order correction (\ref{Jarzynski_like_relation}) are not only effctive near vanshing 
	interaction regime, but can also play a significant role in strong coupling circumstances where $\omega_\mathrm{c}\ge{0.2}~\mathrm{GHz}$ is 
	comparable to $1~\mathrm{GHz}$ namely to $\omega$. Significance of the correction $e^{-\beta\Delta{E}_\mathrm{I}}$ to Jarzynski equaltiy in 
	strong coupling regime can be prominent at least in the specific model we studied numerically here. This agrees with the small probability of 
	significant deviation from (\ref{Crooks_like_relation}) as shown in Fig.~\ref{CR_1st_correction_dev_sig}.

%=================================================================================================================================================
% Disscussion and outlook.
%=================================================================================================================================================
\emph{Disscussion and outlook.}---The perturbative analysis of Crooks relation and Jarzynski equaltiy in the limit of weak coupling shade some 
	light on the study of fluctuation theories in an open system from a new angle. Our results shows that energy trapped in interaction between 
	system and environement is a prominent cause of deviation of fluctuation relations from those established in closed systems as has been shown
	in (\ref{Crooks_like_relation}) and (\ref{Jarzynski_like_relation}). Contractively, heat transfered to (or from) environment seems a much less
	important cause of such a deviation according to (\ref{CrooksRel_E_Q}). Our numerical results provide a possibility that correction with the 
	interactive energy integrated can play a prominent role even in strong coupling regime as in the case we reported in this Letter. It is an 
	interesting question whether in strong coupling regime such a correction would be prominent in general? If not, to what condition, would the 
	effect of this correction extent to the strong coupling regime to a considerable degree? Further, our analysis 
	(\ref{F_tot_1st-order_correction}) indicate that (\ref{1st_law}) should be a good definition of work in open system weakly coupled to a heat 
	bath. It is  consistent with the work definition in a closed system. The result refreshes our understanding of work in quantum regime. 
%=======================================================================================================================================
%	Acknowledge
%=======================================================================================================================================
\begin{acknowledgments}
	This work was supported by Ministry of Science and Technology of China (Grants No. 2016YFA0302104 and 2016YFA0300600), 
	National Natural Science Foundation of China (Grant No. 91536108) and Chinese Academy of Sciences (Grants No. XDB01010000 and XDB21030300).
\end{acknowledgments}
\bibliography{Bibliography.bib}
\end{document}